\documentstyle[12pt,fleqn]{article}
\def\la{\langle}
\def\ra{\rangle}
\def\beeq{\begin{equation}}
\def\eneq{\end{equation}}
\def\beeqa{\begin{eqnarray}}
\def\eneqa{\end{eqnarray}}

\addtocounter{section}{-1}
\begin{document}

\begin{center}

\vspace{2cm}

{\large {\bf {
Electronic Properties of Topological Materials:\\
Optical Excitations in M\"{o}bius Conjugated Polymers
} } }

\vspace{1cm}

{\rm Kikuo Harigaya\footnote[1]{E-mail address: 
\verb+k.harigaya@aist.go.jp+; URL: 
\verb+http://staff.aist.go.jp/k.harigaya/+}}

\vspace{1cm}

{\sl Nanotechnology Research Institute, AIST, 
Tsukuba 305-8568, Japan}\footnote[2]{Corresponding address}\\
{\sl Synthetic Nano-Function Materials Project, 
AIST, Tsukuba 305-8568, Japan}

\vspace{1cm}

(Received~~~~~~~~~~~~~~~~~~~~~~~~~~~~~~~~~~~)
\end{center}

\vspace{1cm}

\noindent
{\bf Abstract}\\
Electronic structures and optical excitations in M\"{o}bius 
conjugated polymers are studied theoretically.  Periodic and 
M\"{o}bius boundary conditions are applied to the tight 
binding model of poly({\sl para}-phenylene), taking exciton 
effects into account.  We discuss that oligomers with a few 
structural units are more effective than polymers for
observations of effects of discrete wave numbers that are
shifted by the change in boundary condition.  Next, calculations of  
optical absorption spectra are reported.  Certain components 
of optical absorption for an electric field perpendicular 
to the polymer axis mix with absorption spectra for an 
electric field parallel to the polymer axis.  Therefore, 
the polarization dependences of an electric field of light
enable us to detect whether conjugated polymers have the 
M\"{o}bius boundary.

\vspace{1cm}

\pagebreak

Recently, low-dimensional materials with peculiar boundary 
conditions, {\sl i.e.}, M\"{o}bius boundaries have been synthesized: 
they are NbSe$_3$ (ref. 1) and aromatic hydrocarbons$^{2)}$, for example. 
A M\"{o}bius strip consists of one surface that does not have the difference 
between the outer and inner surfaces.  The orbitals of electrons 
are twisted while traveling along the strip axis, and 
electronic states can be treated with antiperiodic boundary 
conditions mathematically.  Even though the presence of twisted 
$\pi$-electron systems has been predicted theoretically about 
forty years ago,$^{3)}$ structural perturbations due to topological 
characters might result in physical properties, thus intensive 
investigations are being performed experimentally and 
theoretically.

In a theoretical viewpoint, we have studied boundary condition 
effects in nanographite systems where carbon atoms are arrayed 
in a one-dimensional shape with zigzag edges.$^{4-6)}$  Due to the 
presence of a M\"{o}bius boundary, magnetic domain wall 
states and helical magnetic orders are realized in spin alignments, 
and a domain wall also appears in charge density wave states.  
Such abundant properties have been experimentally observed by 
their unique magnetic properties.$^{7,8)}$

In this paper, we investigate another candidate for twisted 
$\pi$-electron systems: conjugated polymers.  We choose 
poly({\sl para}-phenylene) (PPP) as a model material of conjugated 
polymers.  The structure of PPP is shown in Fig. 1.
Optical excitations in periodic systems have been previously
studied using a tight binding model with long-range Coulomb 
interactions for poly({\sl para}-phenylenevinylene) (PPV),$^{9)}$
PPP, and so forth.$^{10)}$  Exciton effects have been taken 
into account by the configuration interaction method.  Here, 
we study optical excitations in M\"{o}bius PPP strips.  
We expect that such predictive studies of electronic
properties will promote the synthesis of materials with unique
geometries.

We will discuss that oligomers with a few structural units 
are more effective than polymers for measurements of effects 
of discrete wave numbers that are shifted from those of the 
periodic boundary by the M\"{o}bius boundary.  When 
ring-shaped oligomers$^{11,12)}$ are converted into M\"{o}bius
oligomers, such effects could be observed by experiments.
Next we will calculate optical absorption spectra for
the periodic [Fig. 2(a)], M\"{o}bius [Figs. 2(b) and 2(c)]
PPP strips.  We consider two cases of geometries for the M\"{o}bius 
strips: (I) the twist of the bonds is spatially uniform,
as shown schematically in Fig. 2(b), and (II) twist 
positions are spatially limited with respect to the dimensions of
the entire strip, as shown in Fig. 2(c).  We will reveal
that certain components of optical absorption for 
an electric field perpendicular to the polymer axis mix 
with absorption spectra for an electric field parallel 
to the polymer axis.  The polarization dependences of 
an electric field of light enable us to detect whether conjugated 
polymers have the M\"{o}bius boundary.  The 
dependences on the geometries of the M\"{o}bius strips 
are also reported.

We consider optical excitations in PPP using a model$^{9)}$
with electron-phonon and electron-electron interactions.  
The model is shown below:
\beeqa
H &=& H_{\rm pol} + H_{\rm int}, \\
H_{\rm pol} &=& 
- \sum_{\la i,j \ra,\sigma} ( t - \alpha y_{i,j} )
( c_{i,\sigma}^\dagger c_{j,\sigma} + {\rm h.c.} )
+ \frac{K}{2} \sum_{\la i,j \ra} y_{i,j}^2, \\
H_{\rm int} &=& U \sum_{i} 
(c_{i,\uparrow}^\dagger c_{i,\uparrow} - \frac{n_{\rm el}}{2})
(c_{i,\downarrow}^\dagger c_{i,\downarrow} 
- \frac{n_{\rm el}}{2}) \nonumber \\
&+& \sum_{i,j} W(r_{i,j}) 
(\sum_\sigma c_{i,\sigma}^\dagger c_{i,\sigma} - n_{\rm el})
(\sum_\tau c_{j,\tau}^\dagger c_{j,\tau} - n_{\rm el}).
\eneqa
In eq. (1), the first term $H_{\rm pol}$ is the tight 
binding model along the polymer backbone with electron-phonon 
interactions, which couple $\pi$-electrons with modulation 
modes of bond lengths, and the second term $H_{\rm int}$ 
is the Coulomb interaction potential among $\pi$-electrons.  
In eq. (2), $t$ ($> 0$) is the hopping integral between 
the nearest neighbor $i$th and $j$th 
sites in an ideal system without bond alternations; $\alpha$ 
is the electron-phonon coupling constant that modulates 
the hopping integral linearly with respect to the bond 
variable $y_{i,j}$ that measures the magnitude of the 
alternation of the bond $\la i,j \ra$; $y_{i,j} > 0$ 
for long bonds and $y_{i,j} < 0$ for short bonds (the 
average of $y_{i,j}$ is taken to be zero); $K$ is the 
harmonic spring constant for $y_{i,j}$; and the sum is 
taken over pairs of neighboring atoms.  Equation (3) indicates 
the Coulomb interaction among electrons.  Here, $n_{\rm el}$ 
is the average number of electrons per site; $r_{i,j}$ is 
the distance between the $i$th and $j$th sites; and 
$W(r) = 1/\sqrt{(1/U)^2 + (r/a V)^2}$
is the parametrized Ohno potential.  The quantity $W(0) = U$ 
is the strength of the on-site interaction; $V$ is the 
strength of the long-range part ($W(r) \sim aV/r$ in the limit 
$r \gg a$); and $a$ is the mean bond length.

Electron-phonon interactions are treated by classical
approximation with complete lattice relaxation as in ref. 9.
Excitation wave functions of an electron-hole pair are 
calculated by the Hartree-Fock approximation followed 
by the single-excitation configuration interaction method.  
This method, which is appropriate for cases of 
moderate Coulomb interactions, that is, strengths between 
negligible and strong Coulomb interactions, is 
known as the intermediate exciton theory used in ref. 9.
We write singlet electron-hole excitations as
$|\mu, \lambda \rangle = (1/\sqrt{2})
(c_{\mu, \uparrow}^\dagger c_{\lambda, \uparrow} 
- c_{\mu, \downarrow}^\dagger c_{\lambda, \downarrow} )
| g \rangle,$
where $\mu$ and $\lambda$ are unoccupied and occupied states, 
respectively, and $| g \rangle$ is the Hartree-Fock ground state.  
The general expression of the $\kappa$th optical excitation is
$| \kappa \rangle = \sum_{(\mu,\lambda)} D_{\kappa,(\mu,\lambda)}
| \mu, \lambda \rangle$.
The optical absorption spectrum $\alpha(\omega)$ is calculated 
using these excitations for $l = x, y, z$ polarizations:
$\alpha(\omega) = \sum_\kappa E_\kappa L(\omega - E_\kappa)
\langle g | l | \kappa \rangle
\langle \kappa | l | g \rangle$,
where $E_\kappa$ is the optical excitation energy of the
state $|\kappa \rangle$ and $L(\omega)$ is the 
Lorentzian function $L(\omega)=\gamma/[\pi(\omega^2+\gamma^2)]$.

The difference between the periodic and M\"{o}bius boundary 
conditions is regarded as due to the addition of  
antiperiodicity.  As well known in the textbook of condensed 
matter physics, the allowed wave numbers are different 
between periodic and antiperiodic boundary conditions.  
Because the allowed states in the wave number space are 
densely populated in a sufficiently long polymer, the effects 
of the boundary condition difference are too small to 
be observed.  However, in oligomers with a few phenyl 
rings, the allowed states are so sparsely distributed 
that the boundary condition effects could be measured.  
The recent synthesis of M\"{o}bius aromatic systems$^{2)}$
will promote the synthesis of M\"{o}bius polymers in view 
of the presence of many types of ring polymers.$^{11,12)}$

In the following, we discuss the selection rules
for the PPP case.  The electronic states in the wave number 
space are shown for the number of phenyls $N=5$ in Fig. 3. 
The electronic states of the tight binding model with the 
hopping integral $t$ only are shown.  The allowed states for 
the periodic and antiperiodic boundary conditions are 
shown by filled and open circles, respectively.
The allowed wave numbers are $k = j(2\pi/bN)$ for $j=0$ to
$N-1$ in the extended zone scheme for the periodic boundary condition,
$b$ being the size of the unit cell.  They are $k = (j + 1/2)(2\pi/bN)$ 
for $j=0$ to $N-1$ in the M\"{o}bius (antiperiodic) boundary
condition.  If $N$ is of the orders of a few, the allowed wave
numbers are sparsely distributed.  The change in the lowest 
optical transition energy around the $\Gamma$ point between 
the two boundary conditions is expected.  However, the 
distribution of the allowed wave numbers becomes dense as 
$N$ increases.  Therefore, the experimental synthesis of the
M\"{o}bius oligomers with a few phenyl rings will be
necessary for measuring the change of electronic 
structures from those of the periodic oligomers.

First, we discuss the optical absorption of PPP
for the periodic boundary condition.  As in ref. 9, we use the 
following parameters: $\alpha = 2.59$ $t$/\AA, $K=26.6$ $t$/\AA$^2$, 
$U=2.5t$, and $V=1.3t$, where the parameters of the dimensions 
of energies are shown in the unit of $t$.  The $t$ actual value
is 2.3 eV for PPV (ref. 9).  Similar magnitudes are also expected 
for PPP.  The number of phenyl rings is $N=20$.  Figure 2(a) 
shows the geometry of the strip with a periodic PPP chain.
The circle of the polymer axis is assumed to be placed 
within the $x$-$y$ plane.  The $z$-axis is perpendicular to 
the polymer axis.  Figure 4 shows the calculated absorption 
with the Lorentzian broadening $\gamma=0.15t$.  The unit
of the vertical axis is taken to be arbitrary.

When an electric field of light is parallel to the polymer 
axis [Fig. 4(a)], the lowest optical excitations appear at 
an energy of about $1.4t$.  They are excitations between the 
extended states, which are the states at the top of the valence 
band and those at the bottom of the conduction band in Fig. 3.
Off course, the excitation energies calculated by taking
exciton effects into account are different from 
those in Fig. 3, where the electronic states of the
free electron model are displayed.
The higher feature at $2.4t$ is due to the 
optical excitations between the localized occupied and 
unoccupied orbitals.  The localized orbitals are shown in
Fig. 3.  When the electric field is perpendicular 
to the polymer axis [Fig. 4(b)], the optical excitations 
appear at energies larger than $2.2t$.  They are 
excitations between the extended states and the localized 
orbitals.  Figure 4(c) shows the averaged spectrum with 
respect to the polarization of light, {\sl i.e.}, the sum 
over $x$-, $y$-, and $z$-polarizations.  The overall 
features have been observed experimentally.  For details, 
refer to previously published reports.$^{9,10)}$  In energies
higher than $\sim 3t$, optical excitations including
$\sigma$-electrons mix with absorption spectra.  Such  
effects require calculations with both $\pi$- and
$\sigma$-electrons.

Next, we consider two M\"{o}bius boundary cases.  The first 
case (I) is that ring torsions uniformly occur over 
a polymer with the torsion angle $\Psi=180^\circ/N$.  
Phenyl rings helically rotate along with the polymer. 
Figure 2(b) shows the geometry of the polymer strip.
The polymer axis is represented by a circle in the $x$-$y$ 
plane as in the periodic boundary case.  The plane of the
PPP strip is along the $x$-axis where $x>0$ and the
polymer crosses the $x$-axis, and
the polymer plane is perpendicular to the $x$-axis where
$x<0$.  Where the polymer strip crosses the $y$-axis, 
the polymer plane leans by $45^\circ$ within the $y$-$z$ plane.
The second case (II) is that the twist due to the M\"{o}bius 
boundary is localized among five phenyl rings where the polymer 
circle crosses the $x$-axis.  The geometry is shown in
Fig. 2(c).  The twist angle from the torsion is taken  
to be $\Psi=30^\circ$ as a representative value.  
The modulation of the hopping integral
by the torsion is taken into account by the replacement
$t-\alpha y \Rightarrow (t-\alpha y) {\rm cos} \Psi$,
at the bonds where torsions are present.  Hereafter, the 
calculated absorption spectra are shown for the number 
of phenyl rings $N=20$.  The uniform torsion of about $23^\circ$
is not considered, because the modulation of the
overall shape of the optical spectra is small even if we 
consider the uniform torsion to be zero.$^{10)}$

In case I, the plane that includes the polymer is 
almost parallel to the $x$-$y$ plane in a certain part 
of the polymer.  This occurs near the region where the 
polymer crosses the positive part of the $x$-axis 
[Fig. 2(b)].  In this part, an almost perpendicular 
polarization is realized, and therefore mixing of the 
absorption due to this perpendicular polarization 
[Fig. 4(b)] is expected.  Figure 5(a) shows the actual 
calculation.  There are two weak features due to the mixing 
among the energies $2.2t$ and $3.0t$.  There 
is no difference between the $x$- and $y$-polarizations, 
because the system is uniform.  On the other hand, 
the absorption of the perpendicular polarization does 
not show any polarization dependence [Fig. 5(b)].

In case II, the polymer plane can become parallel 
to the $x$-$y$ plane at the region where the twist is present.
This is the part where the polymer strip meets 
the positive region of the $x$-axis [Fig. 2(c)].  
There is a nearly perpendicular polarization in this region.  
The mixing of the perpendicular polarization occurs among 
the energies $3.0t$ as shown in Fig. 6(a), where the 
electric field of light is parallel to the $x$-axis.  
For the case that the field is along the $y$-axis, the 
mixing of the perpendicular polarization is very weak 
because the presence of twists is spatially limited 
in the entire geometry of the polymer chain [Fig. 6(b)].

For the fixed helical torsion angle $\Psi=30^\circ$
and $N=6$, cases I and II are equivalent.
As $N$ increases, differences between the two M\"{o}bius
structures are expected to be more apparent.  Therefore, the selection
rule difference in the wave number space will be 
easier to be observed for short oligomers.  
On the other hand, structure-dependent optical absorption 
will reflect what types of M\"{o}bius rings are formed 
in more long polymers or oligomers ($N > 10$).

In summary, electronic structures and optical excitations in 
PPP with periodic and M\"{o}bius 
boundaries have been studied by taking exciton effects
into account.  In the calculated optical absorption 
spectra, certain components of optical absorption for 
an electric field perpendicular to the polymer axis mix 
with absorption spectra for an electric field parallel 
to the polymer axis.  The polarization dependences of 
an electric field of light enables us to detect whether 
conjugated polymers have the M\"{o}bius boundary.
Therefore, the experimental synthesis of materials with unique
geometries is expected.

\begin{flushleft}
{\bf Acknowledgments}
\end{flushleft}

This work has been supported partly by Special Coordination 
Funds for Promoting Science and Technology, and by NEDO 
under the Nanotechnology Program.

\pagebreak
\begin{flushleft}
{\bf References}
\end{flushleft}

\noindent
1) S. Tanda {\sl et al.}: Nature {\bf 417} (2002) 397.\\
2) D. Ajami {\sl et al.}: Nature {\bf 426} (2003) 819.\\
3) E. Heilbronner: Tetrahedron Lett. {\bf 5} (1964) 1923.\\
4) K. Wakabayashi and K. Harigaya:
J. Phys. Soc. Jpn. {\bf 72} (2003) 998.\\
5) A. Yamashiro, Y. Shimoi, K. Harigaya and K. Wakabayashi:
Phys. Rev. B {\bf 68} (2003) 193410.\\
6) A. Yamashiro, Y. Shimoi, K. Harigaya and K. Wakabayashi:
Physica E {\bf 22} (2004) 688.\\
7) Y. Shibayama {\sl et al.}: Phys. Rev. Lett. {\bf 84} (2000) 1744.\\
8) H. Sato {\sl et al.}: Solid State Commun. {\bf 125} (2003) 641.\\
9) K. Harigaya: J. Phys. Soc. Jpn. {\bf 66} (1997) 1272.\\
10) K. Harigaya: J. Phys.: Condens. Matter {\bf 10} (1998) 7679.\\
11) E. Mena-Osteritz: Adv. Mater. {\bf 14} (2002) 609.\\
12) M. Mayor and C. Didschies: Angew. Chem. Int. Ed. {\bf 42} (2003) 3176.\\

\pagebreak

\begin{flushleft}
{\bf Figure Captions}
\end{flushleft}

\mbox{}

\noindent
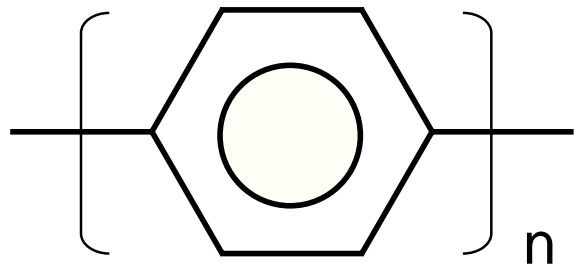
Fig. 1.  Polymer structures of poly({\sl para}-phenylene) (PPP).

\mbox{}

\noindent
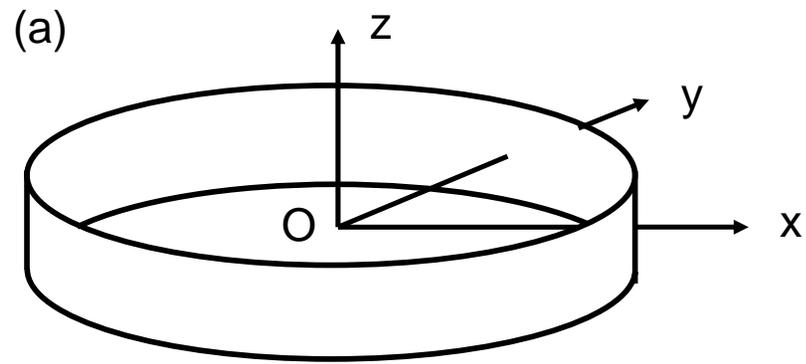
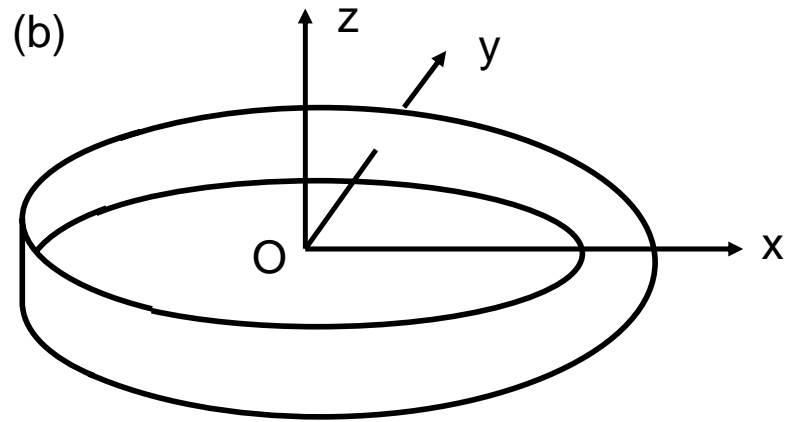
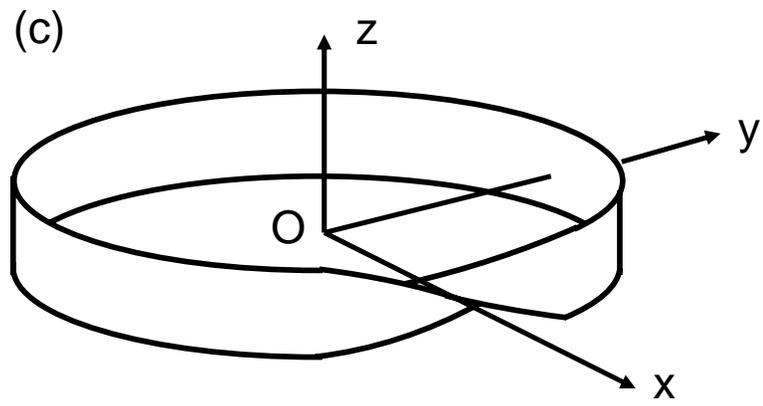
Fig. 2.  Geometries of PPP strips used for calculations:
(a) periodic polymer, (b) spatially uniform M\"{o}bius
polymer, and (c) M\"{o}bius polymer with spatially
localized twists.  The PPP chain is represented schematically
by the strip.  The $\pi$-orbitals of carbons extend 
perpendicularly with respect to the strip.

\mbox{}

\noindent
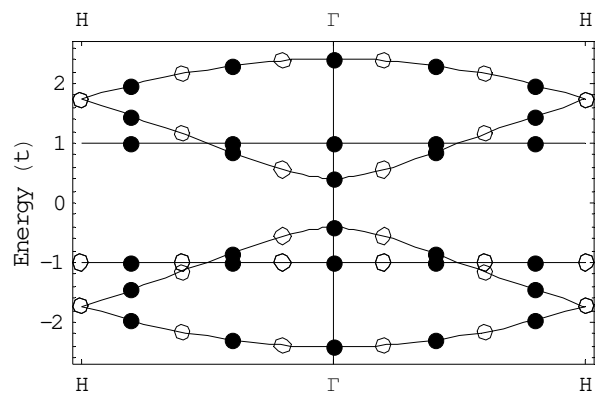
Fig. 3.  Electronic states of PPP in wave number 
space, shown for phenyl ring number $N=5$. The case
of the simple tight binding model with the hopping integral 
$t$ is shown.  The allowed states for the periodic and 
antiperiodic boundary conditions are shown by filled 
and open circles, respectively.  The solid lines represent
the electronic states of the polymer with an infinite length.

\mbox{}

\noindent
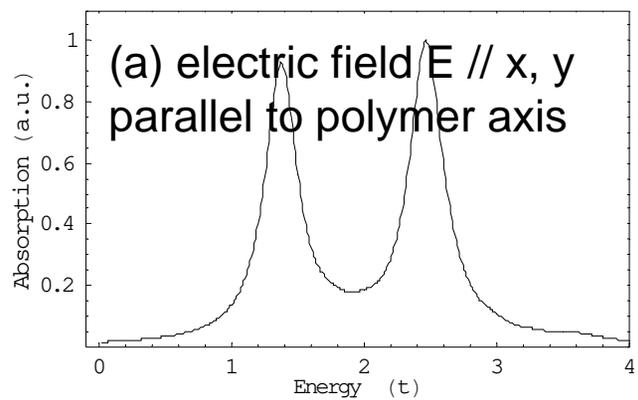
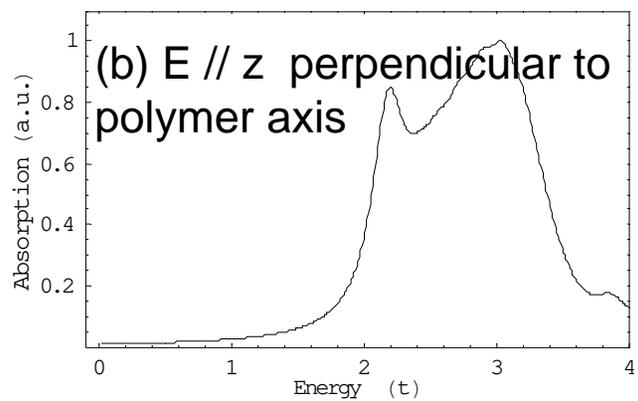
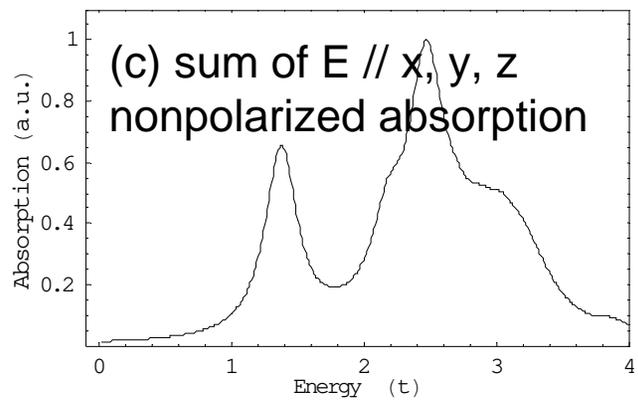
Fig. 4. Optical absorption spectra for periodic boundary case: 
(a) electric field of light parallel to polymer axis, 
(b) electric field perpendicular to polymer axis, and 
(c) nonpolarized absorption. The number of phenyl rings is 
$N=20$.  The vertical axis is shown in arbitrary units,
and the Lorentzian broadening $\gamma = 0.15 t$ is used
in the calculations.

\mbox{}

\noindent
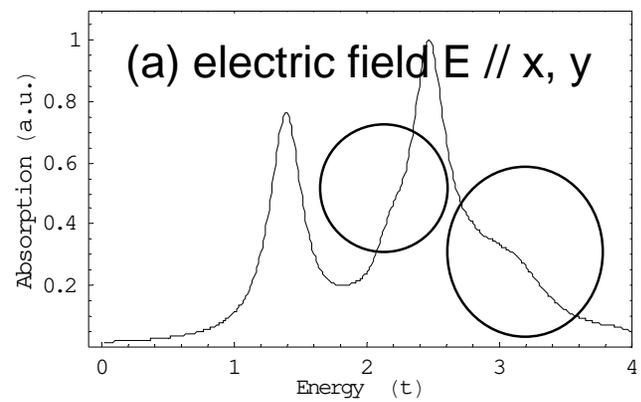
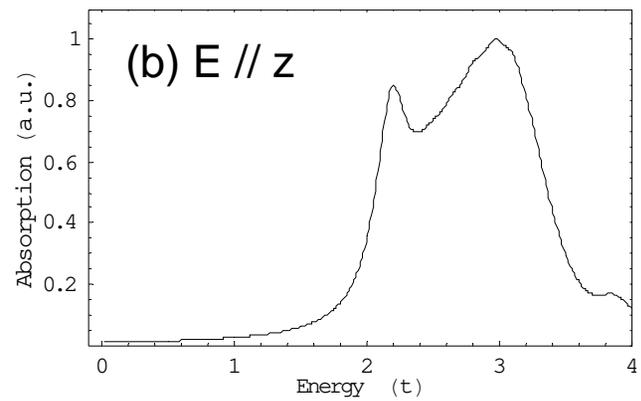
Fig. 5. Optical absorption spectra for M\"{o}bius boundary 
case I: (a) electric field of light parallel to polymer axis 
and (b) electric field perpendicular to polymer axis.  
The circles show features due to the mixing of perpendicular 
polarization.  The number of phenyl rings is 
$N=20$.  The vertical axis is shown in arbitrary units,
and the Lorentzian broadening $\gamma = 0.15 t$ is used
in the calculations.

\mbox{}

\noindent
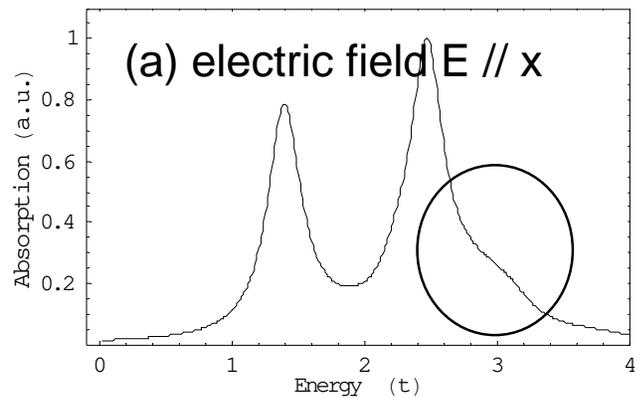
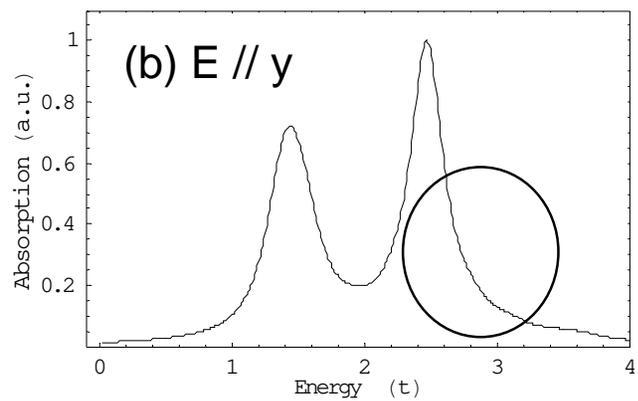
Fig. 6. Optical absorption spectra for M\"{o}bius 
boundary case II: (a) electric field of light parallel 
to $x$-axis and (b)  electric field of light parallel 
to $y$-axis.  The circles show 
features due to the mixing of perpendicular polarization.
The number of phenyl rings is $N=20$.  The vertical 
axis is shown in arbitrary units,
and the Lorentzian broadening $\gamma = 0.15 t$ is used
in the calculations.

\end{document}